\documentclass{article}

\usepackage{PRIMEarxiv}

\usepackage[utf8]{inputenc} 
\usepackage[T1]{fontenc}    
\usepackage[hidelinks]{hyperref}       
\usepackage{url}            
\usepackage{booktabs}       
\usepackage{amsfonts}       
\usepackage{nicefrac}       
\usepackage{microtype}      
\usepackage{lipsum}
\usepackage{fancyhdr}       
\usepackage{graphicx}       
\graphicspath{{media/}}     
\usepackage{amssymb}
\usepackage{mathtools}      
\usepackage{cleveref}           
\usepackage{caption}
\usepackage{subcaption}
\usepackage{multirow}
\usepackage{float}
\usepackage[dvipsnames]{xcolor}
\usepackage{empheq}
\usepackage{soul}

\usepackage{natbib}
\setcitestyle{authoryear,open={(},close={)}}

\begin{document}
\title{Breakdown of Reye's theory 
in nanoscale wear}

\newcommand{\blind}{0}
		\if0\blind
		{
			\author{
			{\Large
			Joaquin Garcia-Suarez $^a$, Tobias Brink $^b$ and 
			Jean-François Molinari $^a$
			}\\
			$^a$Civil Engineering Institute, Materials Science and Engineering Institute,\\
            \'{E}cole Polytechnique F\'{e}d\'{e}rale de Lausanne (EPFL), CH 1015 Lausanne, 
            Switzerland \\
            $^b$ Max-Planck-Institut f\"{u}r Eisenforschung GmbH,\\   Max-Planck-Stra\ss{}e 1, 40237 D\"{u}sseldorf, Germany }
			\date{}
			\maketitle
		} \fi

\begin{abstract}
Building on an analogy to ductile fracture mechanics, we \textcolor{black}{investigate the energetic cost of debris particle creation during adhesive wear}
.  \textcolor{black}{Macroscopically, Reye proposed in 1860 that there is a linear relation between frictional work and wear volume at the macroscopic scale.} Earlier work suggested a linear relation between tangential work and wear debris volume \textcolor{black}{also exists at the scale of a single asperity}, assuming that the debris size is proportional to the micro contact size multiplied by the junction shear strength. However, the present study reveals deviations from linearity \textcolor{black}{at the microscopic scale}. These deviations can be rationalized with fracture mechanics and imply that less work is necessary to generate debris than what was assumed. Here, we postulate that the work needed to detach a wear particle is made of the surface energy expended to create new fracture surfaces, and also of plastic work within a fracture process zone of a given width around the cracks. 
Our theoretical model, validated by molecular dynamics simulations, reveals a super-linear scaling relation between debris volume ($V_d$) and tangential work ($W_t$): $V_d \sim W_t^{3/2}$ in 3D and $V_d \sim W_t^{2}$ in 2D. This study provides a theoretical foundation to estimate the statistical distribution of sizes of fine particles emitted due to adhesive wear processes.
\end{abstract}


\section{Introduction}
\label{Sec:introduction}

Adhesive wear is an unavoidable phenomenon at contacting surfaces subjected to strong adhesive bonds \citep{ref:rabinowicz1995friction,Burwell-Strang}.
It occurs due to microscopic (adhesive) contacts that form wear  particles during sliding \citep{Burwell-Strang,Archard,RABINOWICZ_1958}, which are a result of the surface roughness at small scales \citep{Dieterich-Kilgore, Renard, BT_1939}.
In general, the severity of wear is a function of the applied normal force and is therefore connected to friction \citep{Burwell-Strang,Archard,Rabinowicz_Tabor_Bowden}, but a parameter-free generally applicable model was not yet found \citep{Meng_Ludema, Collins}. 
In light of increasing environmental \citep{EU_reference, brake_review} and health \citep{Kole_et_al} concerns related to fine particle emissions, a mechanistic understanding at the level of single wear particle formation is needed \citep{Vakis, review_numerics}.

Different mechanisms have been put forward to explain adhesive wear. These include wear debris formation with experiments dating back to Archard \citep{Archard} and observed in many cases \citep{ref:bhushan_micronanoscale_1998,ref:chung_fundamental_2003, ref:liu_method_2010, ref:greenwood_deformation_1955,ref:brockley_model_1965}, plastic deformation of contacting asperities, which we may call Holm's mechanism \citep{Holm}, and more recently atom-by-atom attrition \citep{ref:gotsmann_atomistic_2008,ref:bhaskaran_ultralow_2010,ref:sato_real-time_2012,ref:jacobs_nanoscale_2013,ref:stoyanov_nanoscale_2014}. Atomistic simulations \citep{ref:stoyanov_nanoscale_2014,ref:sorensen_simulations_1996,ref:zhong_molecular_2013} generally show plastic deformation, but not Archard's debris formation mechanism. However, recent work \citep{critical_length_scale} could reconcile those observations and revealed that a transition between plastic deformation and debris formation is governed by a critical length scale $d^*$. The parameter $d^*$ describes a minimum junction between contacting asperities for the creation of a wear particle and is related to material properties:

\begin{align} \label{eq:dstar}
    d^* 
    =
    \lambda 
    \frac{\Delta \mathrm{w}}{\tau^2 / 2\mu} \, ,
\end{align}

where $\lambda$ is an order-1 prefactor that encapsulates the influence of the geometry,
$\tau$ is the shear strength, $\mu$ is the material's shear modulus \textcolor{black}{and $\Delta \mathrm{w}$ is the per-crack decohesion work, which is in general equal to the critical energy release rate $G_{Ic}$ for crack opening in mode I, but can be replaced by two times the surface energy $2\gamma_s$ in the case of brittle materials}. The insight provided by this length scale for adhesive wear has been used to elucidate nanoscale friction \citep{Barras2021,ref:brink2021effect}, wear particle formation \citep{PNAS,Frerot2018,Tobias_JMPS}, and surface morphology evolution \citep{Enrico,Milanese2020a}.

The connection between friction and wear has always been intriguing. Archard \citep{Archard} hypothesized that the macroscopic wear volume created during relative sliding of two surfaces is a linear function of the normal force (similar to the tangential force in Coulomb friction) times the sliding distance and inversely proportional to the material's hardness. Reye \citep{Reye} postulated a linear relation between the tangential work $W_t$ and the total wear volume at the macroscopic level, which is also often found in experiments \citep{ref:rabinowicz1995friction,Burwell-Strang,Whittaker,Uetz,Fouvry,Fouvry_2}. At the nanoscale, a similar linear relation between a debris particle's volume $V_d$ and the tangential work $W_t$ required to form it was reported \citep{PNAS}. This linearity was anchored on four assumptions:

\begin{enumerate}
    \item[\#1:] Bowder and Tabor's frictional force argument \citep{Tabor&Bowden}: the necessary peak force to overcome the ``frictional force'' arising from a microcontact is approximately equal to the area of the contact times the shear strength of the material, i.e., $F_{max} \approx \tau A_{contact}$. As, by supposition, this contact leads to debris creation, the area of contact can also be interpreted as a cross-section area of the wear particle $A_{contact} \approx A_{debris}$, and thus it can be related to the junction size (characteristic length of the contact patch), $d$. Combining everything: $F_{max} \approx \tau A_{contact} \sim \tau d^2$. 
    \item[\#2:] Effective sliding distance: in order to form a debris particle, the two surfaces must slide relatively over a distance equal or close to the junction size, $S_{eff} \approx d$.
    \item[\#3:] Tangential work: the work necessary to create the particle, $W_t$, is approximately equal to the effective sliding distance times the peak force, $W_t \approx F_{max} S_{eff}$.
    \item[\#4:] The volume of the debris particle is proportional to the junction size: $V_{d} \sim d^3$. Moreover, assuming that the prefactor connecting the two quantities is close to unity, one reaches $V_{d} \approx d^3$.
\end{enumerate}

Thus, mathematically,

\begin{align}
    \int F_t ds
    =
    W_t 
    \approx 
    F_{max} S_{eff} 
    \approx 
    \tau d^3
    \approx 
    \tau V_d
    \, , \label{eq:Vdebris}
\end{align}

where $F_t$ represents the tangential force, leading to the proposed volume estimator $\hat{V}_d = W_t / \tau$. The salient feature of this model is that it predicts a linear scaling between debris volume and frictional work.
However, despite $\hat{V}_d$ comparing satisfactorily to simulations when $d\approx d^*$ \citep{PNAS}, we will show in the following how the performance of this estimate deteriorates as the size of the asperities increases. In particular, we will discuss molecular dynamics simulations results with contact junctions larger than   $d^*$ that reveal super-linear scaling, and will present theoretical arguments to rationalize our observations. \textcolor{black}{This work seeks to provide a more exhaustive description of the wear process at the single asperity level from an energy-balance standpoint}, leveraging on notions of plasticity and ductile fracture mechanics.

The text is structured as follows. In \Cref{Sec:methods}, we detail the numerical models that enabled the simulation results presented in \Cref{Sec:results}. These results show a super-linear scaling of debris size with frictional work. \Cref{Sec:framework} outlines a new framework to resolve the contradiction with the previous theory that argued for a linear scaling. Further implications are commented in \Cref{Sec:discussion}, while \Cref{Sec:conclusion} presents the final conclusions.  

\section{Methods}
\label{Sec:methods}

\subsection{Numerical simulations}

\begin{figure}
\centering
\captionsetup[subfigure]{justification=centering}
\begin{subfigure}[b]{.45\linewidth}
\includegraphics[width=\linewidth]{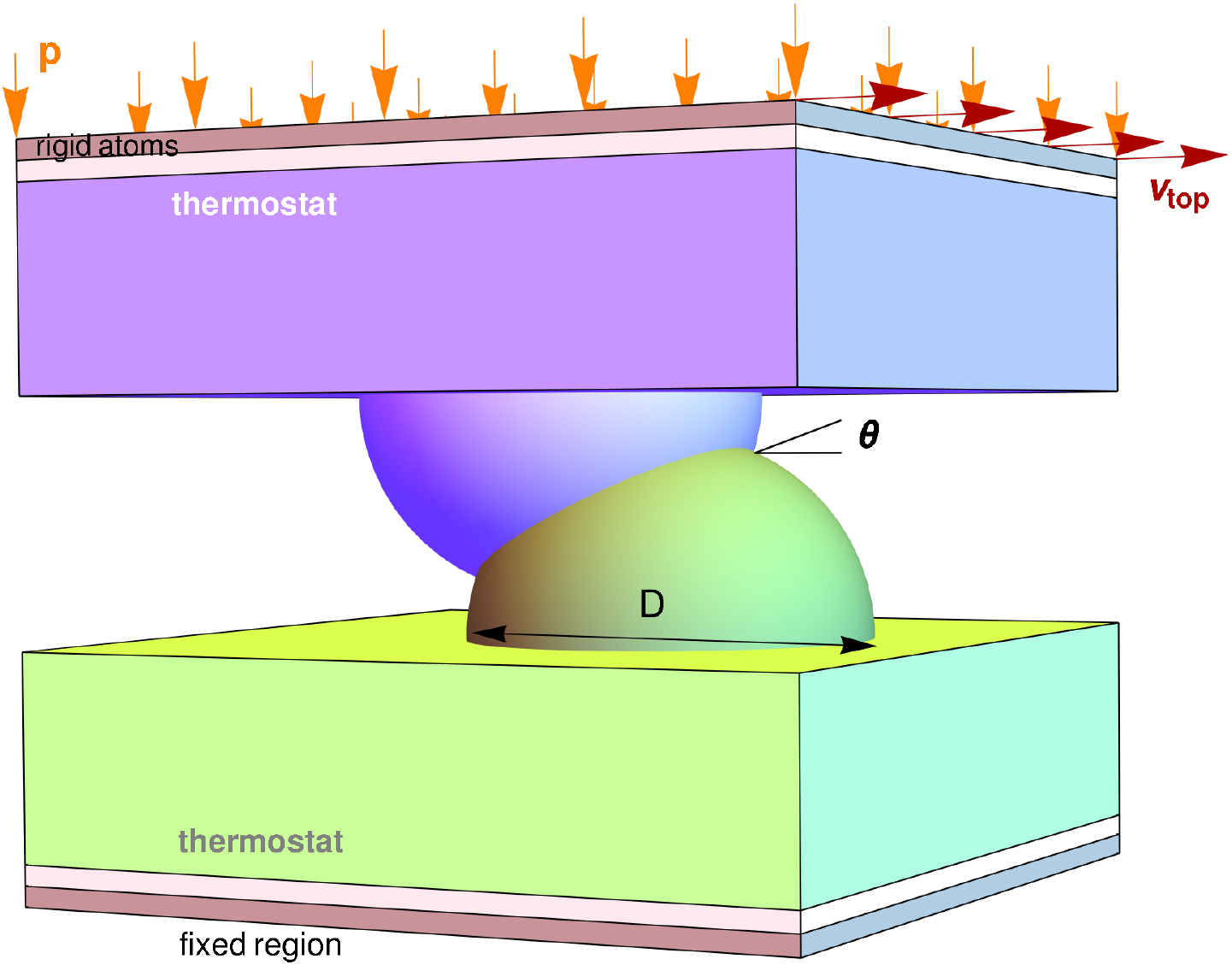}
\caption{\Large (a)}\label{fig:1a}
\end{subfigure}
\begin{subfigure}[b]{.45\linewidth}
\includegraphics[width=\linewidth]{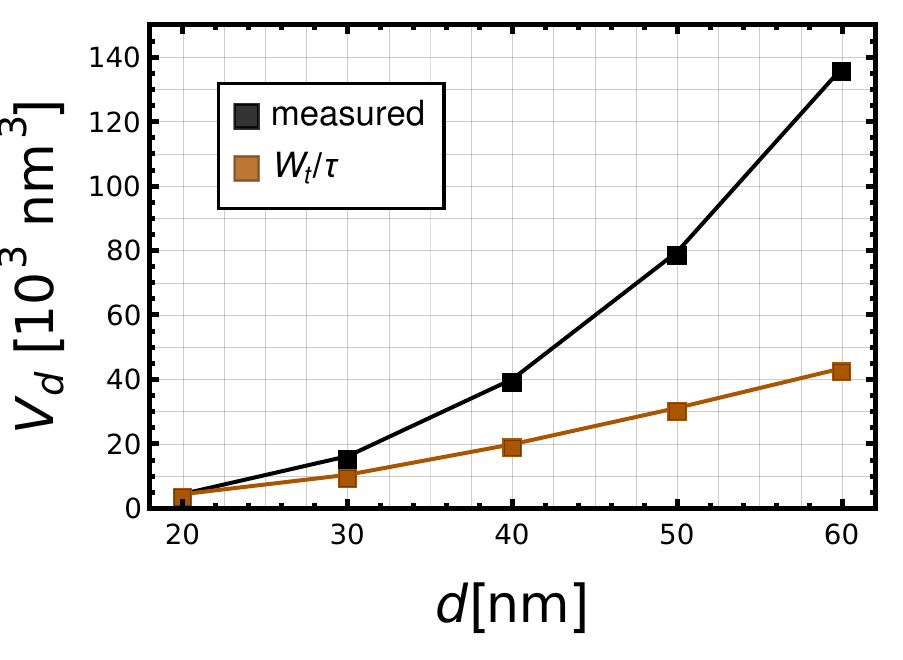}
\caption{\Large (b)}\label{fig:1b}
\end{subfigure}
    \caption{Evidencing the limitations of the debris volume estimate \cref{eq:Vdebris} \citep{PNAS} using both 3D simulations.
    Panel (a): scheme of the 3D simulations, including asperity diameter $D$, contact angle $\theta$ and loading conditions (imposed velocity at top), model presented in \citet{Tobias_PRM}. 
    The two asperities overlap over a circle of diameter $j=0.75d$ (junction size).
    Panel (b): comparison of debris volume measured \textit{in silico} (black) v. estimate (orange), \textcolor{black}{$\theta = 30^{\text{o}}$}. 
    }
    \label{fig:test}
\end{figure}

We performed molecular dynamics simulations on different model asperity geometries in LAMMPS \citep{LAMMPS,Thompson2022}. In three dimensions (3D), we modeled overlapping spherical asperities as described in \citet{Tobias_PRM}. We used a modified Stillinger--Weber \citep{Si_potential} Si potential, with increased bond-angle stiffness \citep{Holland1998erratum}, which better reproduces the fracture behavior at the cost of the other material properties \citep{Holland1998, Holland1998erratum}. The integration time step was $1\,\mathrm{fs}$. In order to obtain an isotropic sample, we produced a glass by melt quenching using the procedure described in Refs.~\citep{Fusco2010, Tobias_PRM}. The material was found earlier \citep{Tobias_PRM} to have a critical length scale of $d^* = 18\,\mathrm{nm}$. Then, the geometry sketched in Fig.~\ref{fig:test}(a) was cut out. Here, we used asperity diameters of $D = 10, 15, 20, 30, 40, 50, 60\,\mathrm{nm}$. The size of the bulk region in the top and bottom crystal was $103 \times 82 \times 15 \,\mathrm{nm}^3$ each. In order to save computational time, we started from asperities that were already in contact on a circular area with diameter $d = 0.75D$. 
Two sets of simulations were considered, one where the contact area between the two asperities is aligned with the sliding direction ($\theta = 0^\circ$) and one where it is inclined ($\theta = 30^\circ$). 
After equilibration for $100\,\mathrm{ps}$ at $T = 300\,\mathrm{K}$, we applied a normal pressure of $0.8\,\mathrm{GPa}$, imposed a tangential displacement velocity of $20\,\mathrm{m/s}$ at the top boundary, and kept the bottom boundary fixed. Langevin thermostats were applied over 4-\AA-thick layers next to the top and bottom boundaries. For the force calculation, the drag force term of the thermostat was subtracted. The tangential force was computed as the reaction force at the top boundary.

We estimate the debris volume by counting the number of atoms in the debris particle and multiplying them by the average atomic volume in the bulk. 
\textcolor{black}{We also estimate the shear strength of the material via independent simulations, finding $\tau = \textcolor{black}{7.9\,\mathrm{GPa} = } 49.31\,\mathrm{eV/nm^3}$.}
Numerical simulations were visualized using OVITO \citep{Ovito} and Mathematica \citep{Mathematica}.

\section{Numerical observations\textcolor{black}{: deviation from linear trend}}
\label{Sec:results}

\begin{figure}
\centering
  \includegraphics[width=\linewidth]{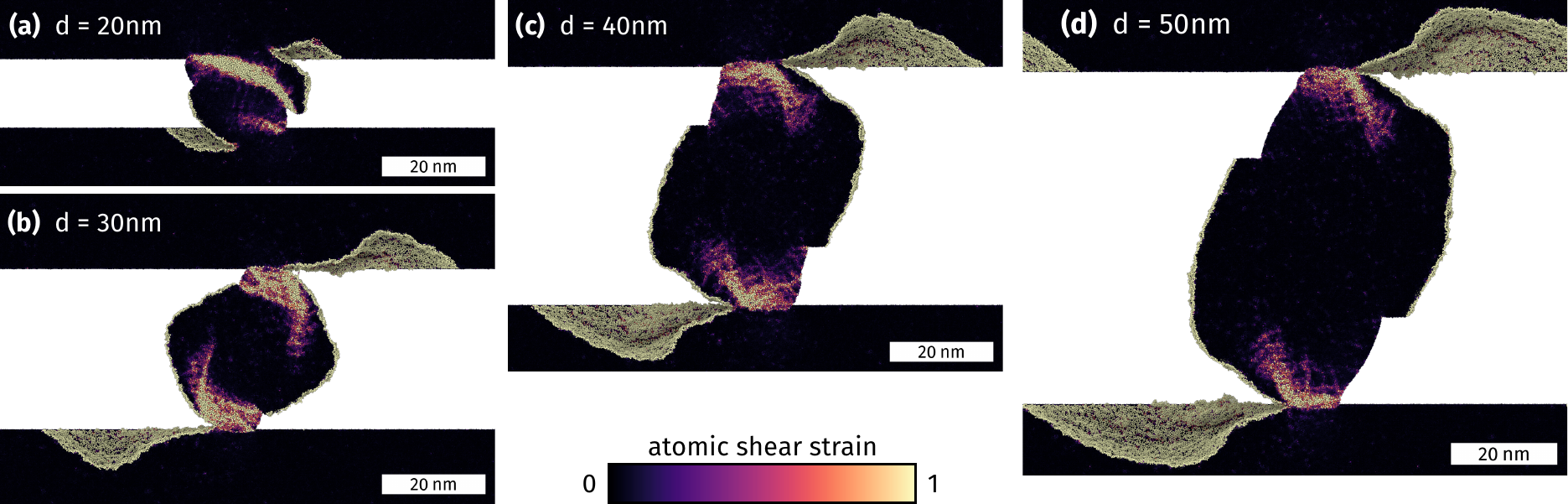}
  \caption{Plastic strain for the 3D simulations with $\theta = 30^\circ$. A slice through the middle of the wear particle is shown to visualize the plasticity in the bulk. (a) For the smallest asperity diameter close to $d^* = 18\,\mathrm{nm}$, significant plasticity occurs in the bulk of the particle. (b)--(d) With increasing asperity size, the plasticity becomes more localized and does not fill the whole particle.}
  \label{fig:Si_images}
\end{figure}

We conducted several sliding simulations of asperity--asperity contacts using the 3D setup. Examples of differently-sized wear particles with $\theta = 30^\circ$ are shown in Fig.~\ref{fig:Si_images}. The critical length scale for this model material was $d^* \approx 18\,\mathrm{nm}$ \citep{Tobias_PRM}, and asperities smaller than this did indeed plastify instead of emitting wear particles (not shown here). We estimated the wear volume for simulations with $d > d^*$ by counting the number of atoms per wear particle and extracted the tangential force from the simulations. Figure~\ref{fig:test}(b) shows the measured wear particle size as a function of the asperity diameter. It can be seen that the volume estimator $\hat{V}_d = W_t/\tau$ from \cref{eq:Vdebris} strongly underestimates the resulting particle volume for large $d$. The quantitative agreement for $d \approx d^*$ is quite good, however. \textcolor{black}{The snapshots in Fig.~\ref{fig:Si_images} already indicate that the relative amount of plastified material decreases and the fracture process becomes more and more brittle, which might suggest that the assumption \#1 of the original model ($F_{max} = \tau A_{contact}$) could be invalid.}

{\color{black}
It is clear that the numerical results are not linear, but proportional to a power of the work with exponent greater than one. This means that the root cause of this disagreement cannot be a faulty estimation of the shear strength $\tau$, but must be associated to the breakdown of one of the four assumptions discussed in \Cref{Sec:introduction}.  
}
 
\section{New framework based on ductile fracture mechanics}
\label{Sec:framework}

The numerical results reveal the necessity to extend the current theory. We note that the details of the crack propagation process are not explicitly accounted for during the derivation of \cref{eq:Vdebris}. 
The theory of Linear-Elastic Fracture Mechanics presupposes that the strain energy within the body goes into breaking bonds between atoms, which in turns means that new surfaces are formed and a crack propagates.
Griffith's original energy balance argument \citep{Griffith} hinges on the assumption that the plastic dissipation occurring due to the stress concentration at the crack tip is a small percentage of the total energy being dissipated. 
To quantify the amount of plasticity accompanying the fracture process, the plastic radius around the crack tip, $r_p$ \citep{Fracture-textbook}, is used. This parameter is traditionally presented as 

\begin{subequations}
\begin{align}
    r_p 
    &=
    {1 \over \pi}
    \left(
    {K_I \over \sigma_{\mathrm{y}}}
    \right)^2 \, ,
    \intertext{where $K_{I}$ is the mode I stress intensity factor, which reaches its critical maximum value at the initiation of crack growth, $K_I = K_{Ic} = \sqrt{G_{Ic}E'}$ where $E' = E$ (Young's modulus) in plane stress and $E' = E/(1-\nu^2)$ ($\nu$ being Poisson's ratio) in plane strain, and $G_{Ic}$ is a material parameter termed ``critical energy release rate'' (in mode I). Thus}
    r_p|_{max}
    &=
    {1 \over \pi}
    {
    G_{Ic} E' 
    \over 
    \sigma_{\mathrm{y}}^2
    } \, ,
\end{align}    
\end{subequations}

defines the characteristic size of this plastic region. 
The condition of small-scale yielding (SSY) sets the range of validity of brittle fracture in terms of the plastic radius and a characteristic geometric length of the system. 
A generally agreed-upon test is checking if $L \ge 25r_p$ \citep{Hutchinson} ($L$ being any of the characteristic geometrical in-plane lengths involved, in this case we can take $L = d$). If so, the use of the brittle approximation is warranted and crack plasticity can be ignored. 
If that is not the case, plasticity can be taken into account by modifying the brittle fracture criterion: in the brittle case, we had $G_{Ic} = 2 \gamma_s$, while in the ductile one $G_{Ic} = 2 (\gamma_s + \gamma_p)$, where the material parameter $\gamma_p$ represents the plastic energy dissipation per unit of new surface area. 

Notice that $r_p$ and $d^*$ scale the same but differ in their prefactors due to the difference in geometry and loading mode (tension versus shear).
In the case of single-asperity wear, $L$ is taken as the junction size $d$, and SSY will not hold when $d / d^*  \sim d / r_p|_{max} \ll 25$.

\begin{figure}
    \centering
    \includegraphics[width=\textwidth]{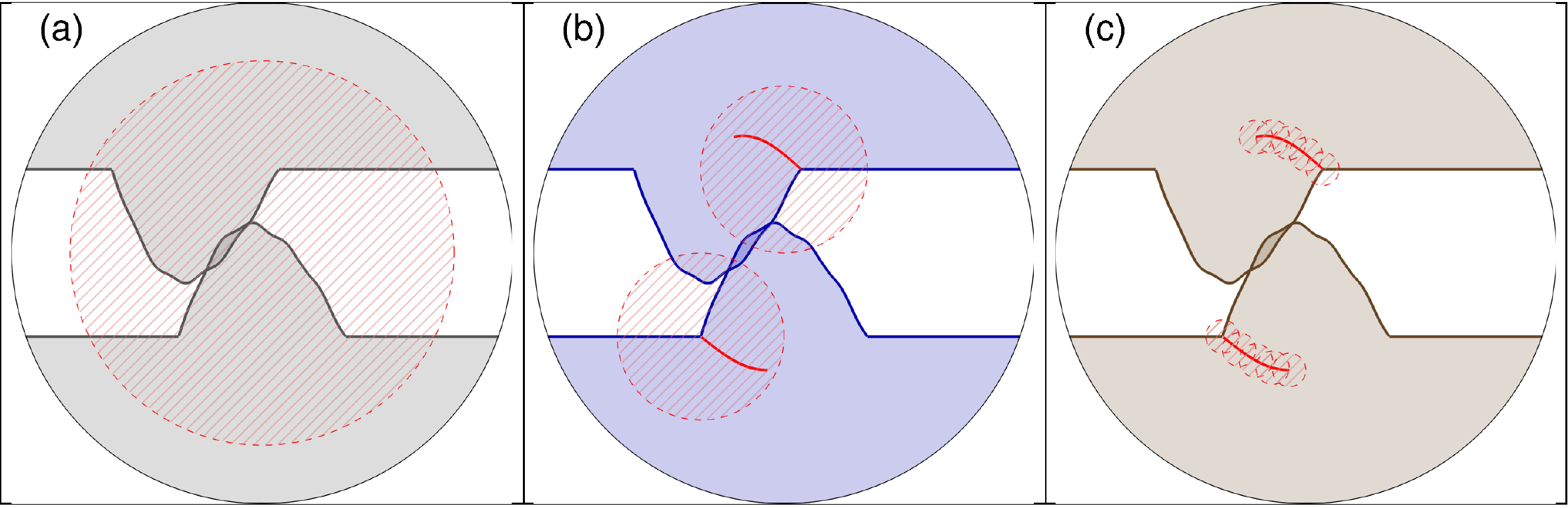}
    \caption{Different behavior for the same system (in terms of scale and geometry) of interlocking asperities, but changing the material. Each color represents a different material with properties yielding different $\ell_{FPZ}$, whose value can be interpreted as the diameter of the patterned red area. (a) System  with $d \ll \ell_{FPZ}$:  all-encompassing plasticity. (b) System  with $d > \ell_{FPZ}$ while $d / \ell_{FPZ} \approx 1$: plasticity widely present, but cracks can nucleate around spots of stress concentration (compare to results from simulations using the phase-field method in \citet{Stella}). (c) System  with $d \gg \ell_{FPZ}$:  localized plasticity around the fracture path (compare to phase-field results from \citet{Reinoso}).}
    \label{fig:material_effect}
\end{figure}

There is yet another parameter that appears in this context \citep{FPZ}, the fracture process zone (FPZ) around a crack tip. It displays the same scaling in terms of the mechanical properties, i.e., $d^* \sim \ell_{FPZ}$ too in the case of brittle materials
:
\begin{align} \label{eq:l_FPZ}
    \ell_{FPZ} 
    \sim
    { \Delta \mathrm{w} 
    \over 
    \tau^2 / \mu} \, .
\end{align}
This parameter represents the size of the damaged region that either eventually nucleates a crack or along which the crack extends; it is characterized by stress concentrations, micro-crack formation and/or other degradation processes. 
The fact that the critical junction size scales in the same fashion as the fracture process zone suggests a reinterpretation of the former: 
stress concentrations around small asperities can only nucleate cracks if the asperity itself is large enough to host a fracture process zone within it, thus $d^*$ can be thought as characterizing the minimal geometry of the system formed by interlocked asperities that can fit a FPZ that nucleates a crack, which leads to third-body formation.
\Cref{fig:material_effect} represents schematically three possible scenarios related to the prior discussion, each one of them arising from changing the material properties while maintaining exactly the same geometry and scale: 
panel (a) corresponds to a material with $\ell_{FPZ} \gg d$ where stress concentrations can only lead to plasticity and surface smoothing (Holm's mechanism), in (b) $\ell_{FPZ} \approx d$ so substantial plasticity gives rise to cracks and eventually leads to third-body creation, and (c) represents a brittle material with $\ell_{FPZ} \ll d$ wherein fractures nucleate before inelastic deformation takes place (but plasticity still appears around the crack path).

\subsection{Discussion of previous work in the context of fracture mechanics}

One of the premises that led to both \cref{eq:dstar} and \cref{eq:Vdebris} is that the volume that is plastified during sliding prior to debris creation, $V_p$, scales proportionally to the volume of the region surrounding the contact patch, and thus is similar to the final particle volume, $V_p \sim V_d \sim d^3$. 
This was verified in simulations in which $d \approx d^*$ \citep{ramin_wear}, where substantial inelasticity can be observed via post-processing or just by looking at the permanent shape changes in the asperities prior to detachment. 
This is not surprising: the transition from a state that is plasticity-dominated to one where fracture also appears does not mean that plasticity is excluded in the latter. Actually, substantial inelastic deformation can accumulate before the cracks grow \citep{ramin_wear} (see intermediate panel in \Cref{fig:material_effect}). 
On the other hand, in the limit of $d \gg d^*$, fractures develop before large deformations and inelasticity can occur.

Resorting again to the 3D simulations
, let us visualize the extent of plastic deformation as the asperity size increases. 
\Cref{fig:Si_images}(a) shows the smallest size ($d \approx d^*$), where plasticity penetrates the bulk of the system formed by the contacting asperities, while in \Cref{fig:Si_images}(d) traces of inelastic activity are only found in a narrow region close to the crack path. 
The latter hints at a ductile fracture, in which substantial plasticity accompanies the crack tip trajectory, the volume of the plastified region at the end of the crack propagation process being $V_p \sim d^2 \ell_{FPZ}$, as the full crack path area is equivalent to the asperity base ($\sim d^2$). \textcolor{black}{Note that this is a simplification, as the actual shape of the plastified region is more complex, but we assume that the scaling of the plastic zone size is still valid.} Therefore, we find that $V_p \sim V_d$ only when $d \sim \ell_{FPZ} \sim d^*$.

In order to understand that, we decompose the tangential work into $W_{plastic}$ (work gone into permanent deformation anywhere in the asperity) and $W_{debonding}$ (work invested in bond breaking). Energy dissipated as heat in other processes can be neglected assuming close to quasi-static loading conditions.
The step-by-step reasoning goes as follows: start from

\begin{subequations}
    \begin{align}
        W_t 
        &\approx 
        W_{plastic} + W_{debonding} \, ,
    \shortintertext{so dividing by $\tau V_{d}$ gives}
        {W_t \over \tau V_{d}}
        & =
        {\hat{V}_{d} \over V_{d}}
        \approx
        {W_{plastic} \over \tau V_d}
        +  
        { W_{debonding} 
        \over 
        \tau V_{d} }\, .
    \end{align}
\end{subequations}

\textcolor{black}{A volume around the crack tip is plastified while the crack propagates. We assume a characteristic local plastic strain $\varepsilon_p$ after which the crack propagates one step further, plastifying a new volume up to a strain $\varepsilon_p$, and so on (cf.~Fig.~\ref{fig:material_effect}(c)). Recall that along the full crack path a volume of $V_p \sim d^2 \ell_{FPZ}$ will be plastified, so we can express the total plastic work as $W_\text{plastic} \approx \tau V_p \varepsilon_p$.} Under the further assumption that $\varepsilon_p = \text{const.}$ (which is likely material dependent), we obtain the scaling

\begin{align}
    {W_{plastic} \over \tau V_d}
    \sim
    { \tau V_p \varepsilon_p
    \over 
    \tau V_d}
    \sim
    {V_p 
    \over 
    V_d}
    \sim 
    { d^2 \ell_{FPZ}
    \over
    d^3}
    \sim 
    {d^*
    \over
    d}
    \, .
\end{align}

The first term thus decays as the size increases, consistent with what we have already seen in \Cref{fig:Si_images}. With the remaining term we find

\begin{subequations}
    \begin{align}
        {W_{debonding} \over \tau V_d}
        &= 
        { 2 \gamma_s A_{created} 
        \over 
        \tau V_d}
        \sim
        { 2 \gamma_s 
        \over 
        \tau d}
        \sim
        \left(
        {\tau 
        \over 
        \mu}
        \right)
        {d^* 
        \over 
        d}
         \, ,
    \shortintertext{where $A_{created}$ is the total new area created by the fractures. The factor inside the parenthesis does not change if the material remains the same, so if only the size changes}
        {W_{debonding} 
        \over 
        \tau V_d}
        & \sim 
        {d^* \over d} \, .
    \end{align}
\end{subequations}

Thus, both parts of the tangential work have the same scaling and we reach

\begin{align}
    {W_t \over \tau V_{d}}
    &=
    {\hat{V}_{d} \over V_{d}}
    =
    \mathcal{O}
    \left(
    {d^* \over d}
    \right)
    \, . \label{eq:scaling_d_dstar}
\end{align}
This result means that the estimate can perform well for the smallest asperities ($d \approx d^* \Rightarrow \hat{V}_{d} \approx V_{d}$), 
but it may underpredict the debris volume as the size increases ($d \gg d^* \Rightarrow \hat{V}_{d} \ll V_{d}$).

\begin{figure}[H]
\centering
\includegraphics[width=\textwidth]{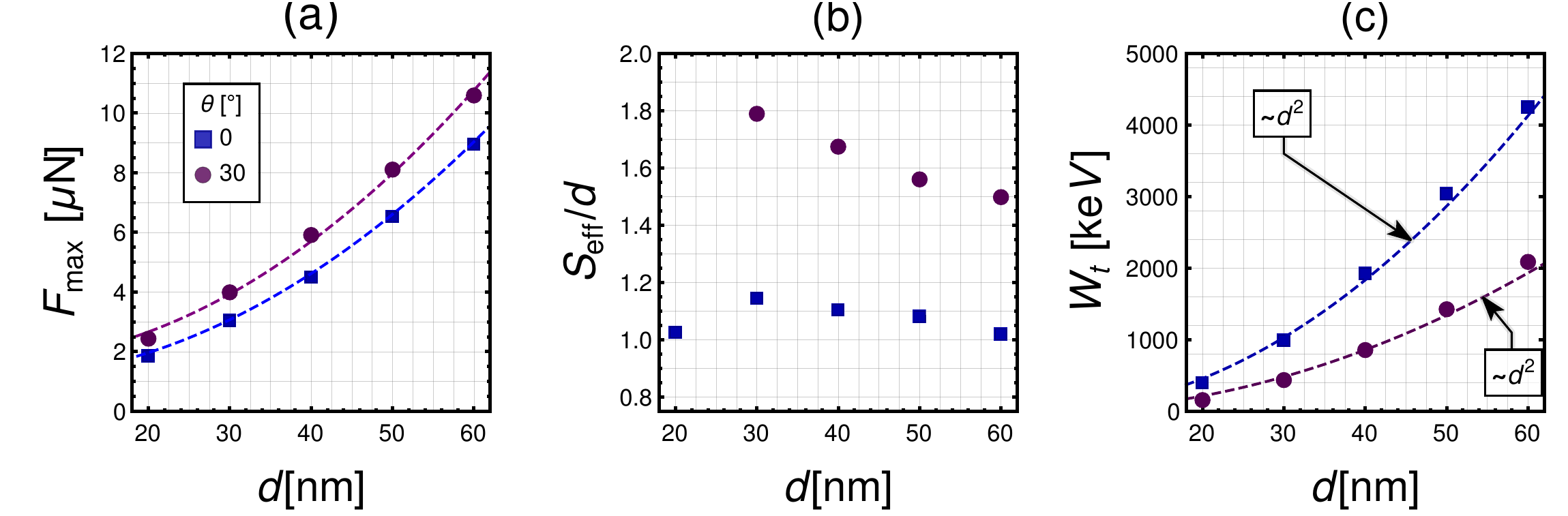}
\caption{\textcolor{black}{Verifying the assumptions behind eq.(2). (a) Assumption \#1: Quadratic scaling of peak force with junction size in 3D: results for two geometries (two contact angles ($\theta$)). (b) Assumption \#2: Decreasing effective sliding as a portion of junction size $d$. (c) Quadratic scaling of tangential work with junction size.}
}
\label{fig:work_scalings}   
\end{figure}



\subsection{Insights from a continuum beam model}
\label{Sec:continuum}

\begin{figure}[H]
    \centering
    \includegraphics[width=\textwidth]{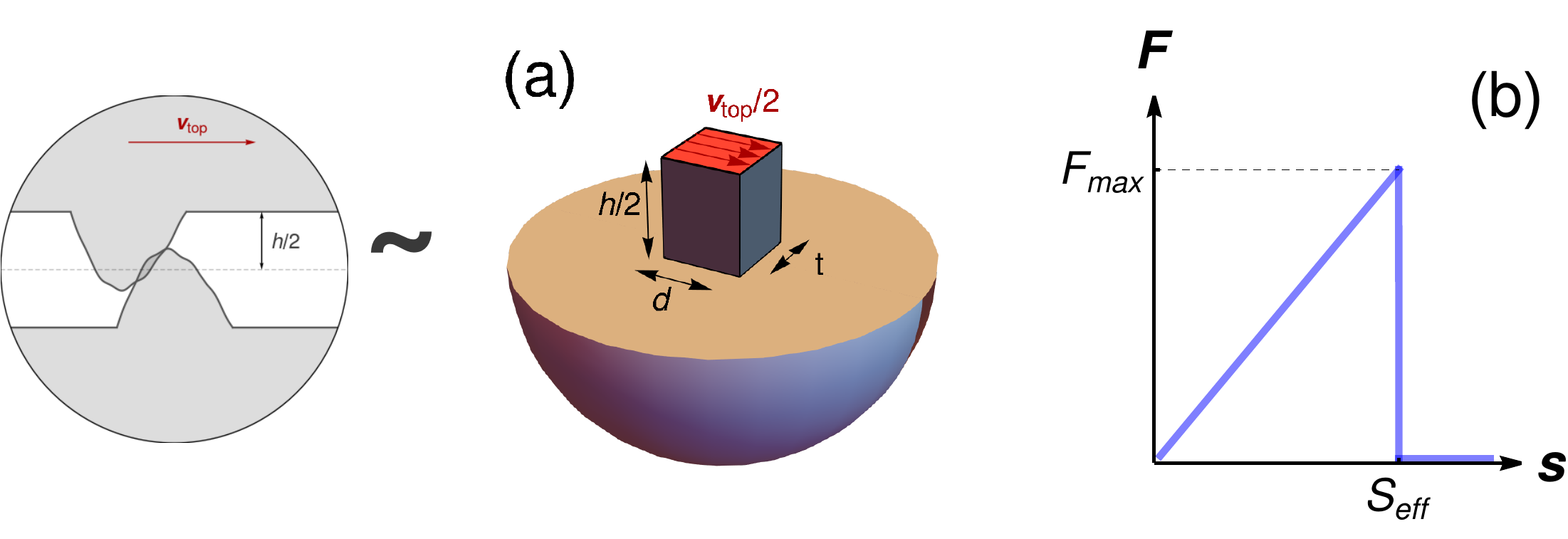}
    \caption{Analytical model scheme. (a) Abstraction from contacting asperities into Timoshenko beam system that can be halved resorting to anti-symmetry conditions \citep{ramin_fracture_modes}. (b) Idealized loading evolution: linear up to point defined by $F_{max}$ and $S_{eff}$, followed by almost-instantaneous drop (in reality, it would have a large slope proportional to the Rayleigh wave velocity). }
    \label{fig:das_beam}
\end{figure}

We supplement the previous discussion with an analytical continuum model presented in \citet{ramin_fracture_modes}, which helps us ascertain the system behavior as its size increases and the continuum limit is approached. 
The idealized model considers two interlocking asperities as a rectangular cross-section Timoshenko beam \citep{Timoshenko} having out-of-plane thickness $\mathrm{t}$, while the other cross-section length corresponds to the junction length $d$. 
The height of each asperity is $h/2$, so the interlocked system has a total height of $h$, see \Cref{fig:das_beam}(a).

Assuming low-velocity, displacement-controlled loading conditions, we grant that most of the elastic energy is released much faster than the time that it takes for the remote loading to change once the failure conditions are attained, i.e., we assume a sliding-force relation as in \Cref{fig:das_beam}(b). This choice will be discussed at the end of the section.

Each asperity can be considered independently exploiting antisymmetry conditions \citep{ramin_fracture_modes}. 
We consider an imposed displacement $u|_{y=h/2}$ instead of an imposed force. 
An estimate of the stiffness $K$ of the system as a function of crack length $a$ was also provided in \citet{ramin_fracture_modes} \textcolor{black}{(the cracks of length $a$ are assumed to grow at the base of each beam)}:

\begin{align}
  \mathrm{K}(a)
  =
  \left[
  { 
  4(h/2)^3
  \over
  E (d-a)^3 \mathrm{t}  
  }
  +
  { h/2 
  \over 
  \kappa \mu (d-a) \mathrm{t}} 
  \right]^{-1}
  \, ,
\end{align}

where $\kappa$ is the shear coefficient \citep{Timoshenko}, \textcolor{black}{this dimensionless parameter appears in Timoshenko theory as a correction factor to properly account for the real distribution of shear stresses in the cross-section}. Note the different scaling of each addend in terms of $h/d$, the aspect ratio of the asperity. 
This approximation assumes that the beams are clamped to a rigid half-space, which is obviously not the case (the connection of the asperities to the surfaces provides extra compliance); however, since we are primarily interested in a scaling analysis, this simplification does not represent an important drawback. 
Using the simple beam model, the total energy of the system is twice the energy stored in each beam:

\begin{align}
    E_{total}
    =
    \mathrm{K}(a)
    u^2|_{y=h/2} \, ,
\end{align}
thus the critical sliding $u = S_{eff}$ that triggers crack propagation can be computed as the displacement at which the critical energy release rate $G_{Ic}$ is attained: 
\begin{align} 
    G
    =
    {1 \over \mathrm{t}}
    {\partial E_{total} 
    \over 
    \partial a}
    =
    -
    {S_{eff}^2 \over \mathrm{t}}
    {\partial \mathrm{K} 
    \over 
    \partial a}
    =
    G_{Ic}
    \, .
\end{align}

If we further assume, on top of low aspect ratio, that the asperities are initially defect-free (so $a/d \to 0$) and that unstable crack growth happens immediately after $G_{Ic}$ is attained, see \Cref{fig:das_beam} panel (b), this yields
\begin{align} \label{eq:S_eff}
    S_{eff}
    =
    \left(
    {
    G_{Ic} \mathrm{t} 
    \over 
    {\partial \mathrm{K} / \partial a}|_{a \to 0}}
    \right)^{1/2}
    =
    2 \sqrt{ G_{Ic} h / \kappa \mu}
    +
    \mathcal{O}
    \left[
    \left(
    {h \over d}
    \right)^2
    \right] \, ,
\end{align}

where the second addend represents terms that will be meaningful only if the assumption of low aspect ratio was removed.

The total tangential work done over the system (two asperities) to the point when sudden failure by unstable crack growth happens is

\begin{align}
    W_t
    =
    \mathrm{K}(0)
    S_{eff}^2
    =
    2 \mathrm{t} d G_{Ic} 
    =
    4 \mathrm{t} d (\gamma_s + \gamma_p) 
    \, ,
\end{align}

that is, the energy necessary to grow two cracks over an area $td$ (the cross-section area) on each asperity. See that if we further assume that $t \sim d$ (the out-of-plane thickness being of the same order as the other cross-section length) we reach $W_t \sim d^2$. This result indicates that the energy invested in sliding leading to debris creation can be controlled by fracture and be proportional to area, instead of being controlled by plasticity and proportional to volume.

It is clear that the second assumption \cref{eq:Vdebris} rests upon ($S_{eff} \approx d$) will enter in conflict with the latter result: 
see the $\sqrt{h}$ proportionality in \cref{eq:S_eff}, i.e., the sliding distance to failure is not a linear function of the size. 
\textcolor{black}{This is consistent with our numerical observations, \Cref{fig:work_scalings} panel (b). Note that the trend does not apply for the smallest sizes ($d = 20 \, \mathrm{nm}$), in which the particle does not fully detach as it sticks to the surfaces (see \Cref{fig:Si_images}(a) and Figure B.5(a) in the supplementary material). For the larger sizes, which do lead to neat third-body formation, the trend is met.}
See that both the numerical and analytical results do yield $S_{eff} \approx d$ for the smallest asperities.

\textcolor{black}{Likewise, \Cref{fig:work_scalings} panel (a) displays the maximum force measured in the simulations, and reveals a quadratic scaling that appears consistent with the Bowden and Tabor model \citep{Tabor&Bowden}, even though the continuum model suggests}

\begin{align}
    F_{max} 
    = 
    \mathrm{K}(0) S_{eff} \approx \sqrt{G_{Ic} h \mu \kappa} \, .
    \label{eq:Fmax}
\end{align}

\textcolor{black}{This indicates that  the maximum force still seems to be dictated by the significant plasticity that can be observed at the simulation sizes studied here (see for example Fig.~\ref{fig:Si_images}) and depends linearly (2D) or quadratically (3D) on the contact size. Nevertheless, the sliding distance $S_{eff}$ is in agreement with the continuum model, which also means that force dropoff after reaching $F_{max}$ must become steeper for larger the asperities.}

\textcolor{black}{In conclusion, given the appraisals obtained from \cref{eq:S_eff,eq:Fmax}, the data shows that the assumption \#2 ($S_{eff} \approx d$) used to derive the original volume estimator in \cref{eq:Vdebris} is violated. We expect from the continuum model that even assumption \#1 ($F_{max} \approx \tau d^2$) would not be followed for much larger asperity--asperity contacts, but the data presented in this paper likely covers a range that is too small to clearly see this transition to the idealized case used to derive \cref{eq:Fmax}.}
These findings also make intuitive sense if we think of crack propagation regimes being a function of the asperity size:

\begin{itemize}
    \item Small junctions, $d \sim d^*$, strength-controlled propagation: the stress intensity factor is strongly influenced by the surroundings' geometry, and its stiffness by extension. 
    The process is driven by the remote sliding condition, and the strain energy stored during prior deformation (along with the stiffness of the system) is lost gradually at a rate proportional to the imposed velocity. 
    \item Large junctions, $d \gg d^*$, toughness-controlled: even though the crack nucleation will depend on the local geometry of the system, most of its growth will take place far from the model's edges, being effectively independent of the geometrical features. The stored energy is released at a fast rate (related to rapid crack propagation), as assumed in \Cref{fig:das_beam}(b), while the crack grows in toughness-controlled conditions.   
\end{itemize}

As the size of the system is increased (greater $d$) from an initial size $d^*$, the relative level of toughness-controlled propagation will in turn also increase, in detriment of ever smaller portions of strength-controlled. We further substantiate this point in the supplementary discussion around \Cref{fig:Son_data}.  

\textcolor{black}{Finally, let us mention that the same trends reported herein are also observed in the 2D simulations, \Cref{fig:assumptions_check}: effective sliding as a portion of asperity size $d$ decreases as the size increases while the peak force does scale linearly as predicted by Bowden and Tabor's model in 2D ($F_{max} \propto d$).}

\subsection{A new \textcolor{black}{scaling relation}}
 
The system approaches a continuum-like situation similar to the one studied in the prior section as the size of the junction increases over the threshold value, hence the energy released at the crack tip becomes the dominant source of dissipation, i.e.,
\begin{align} \label{eq:work_prop_area}
    W_{t}
    \approx
    G_{Ic} A_{created} \, ,
\end{align}
where $A_{created}$ represents the total new area, the sum of the two new surfaces, one in each asperity. This work must be approximately equal to the strain energy stored during sliding, it does not account for the extra work that goes into the next stage: ``rotating out'' the particle.

Extrapolation of the attested scalings $V_d \sim d^3$ (3D) and $A_d \sim d^2$ (2D) \citep{PNAS} combined with \cref{eq:work_prop_area} leads to
\begin{subequations}
\begin{empheq}[left={W_t \approx G_{Ic} A_{created} \sim \empheqlbrace \,}]{align}
      & G_{Ic} d^2 \sim G_{Ic} V_d^{2/3}
    \Rightarrow
    V_d \sim \left( {W_t \over G_{Ic}}
    \right)^{3/2} \text{  in 3D}
    \, , \label{eq:scaling_3D}
    \\
      & G_{Ic} t d \sim G_{Ic} \mathrm{t} A_d^{1/2}
    \Rightarrow
    V_d =\mathrm{t} A_d \sim {1 \over \mathrm{t}} \left( {W_t \over G_{Ic}} \right)^{2}
     \text{  in 2D}
    \, . \label{eq:scaling_2D}
    \end{empheq}
    \label{eq:new_scalings}
\end{subequations}
Here, $\mathrm{t}$ is the thickness of the 2D system. These scaling relations are the main contribution of this article.
The corresponding pre-factors require the estimation of the total crack area, which in turn must depend on the specific geometry of the junction and the asperities (its size and presence of the stress concentration spots). 

{\color{black}
Assuming that both the prefactors (called $\mathrm{k}$ in both cases) and the fracture toughness were known, we would write

\begin{subequations}
    \begin{align}
        \hat{V}_d
        &=
        \mathrm{k}
        \left(
        \frac{W_t}{G_{Ic}}
        \right)^{1.5} \, .
        \label{eq:new_Vd_3D}
    \shortintertext{for the 3D case, and the 2D one}
        \hat{V}_d
        &=
        {\mathrm{k} \over \mathrm{t}}
        \left(
        {W_t \over G_{Ic}}
        \right)^{2} \, ,
        \label{eq:new_Vd_2D}
    \end{align}
\end{subequations}

See how \cref{eq:new_Vd_3D} captures the super-linear scaling of volume with work that we see in the data, \Cref{fig:6d,fig:6b}, and how the estimators can provide quantitative predictions if the unknowns $\mathrm{k}$ and $G_{Ic}$ are chosen accordingly, \Cref{fig:6a,fig:6c}.

In conclusion, a super-linear scaling of debris particle volume with tangential work is observed at the nanoscale as the size of the asperities and the junction patch increases over $d^*$. 

\begin{figure}
\centering
\captionsetup[subfigure]{justification=centering}
\begin{subfigure}[b]{.45\linewidth}
\includegraphics[width=\linewidth]{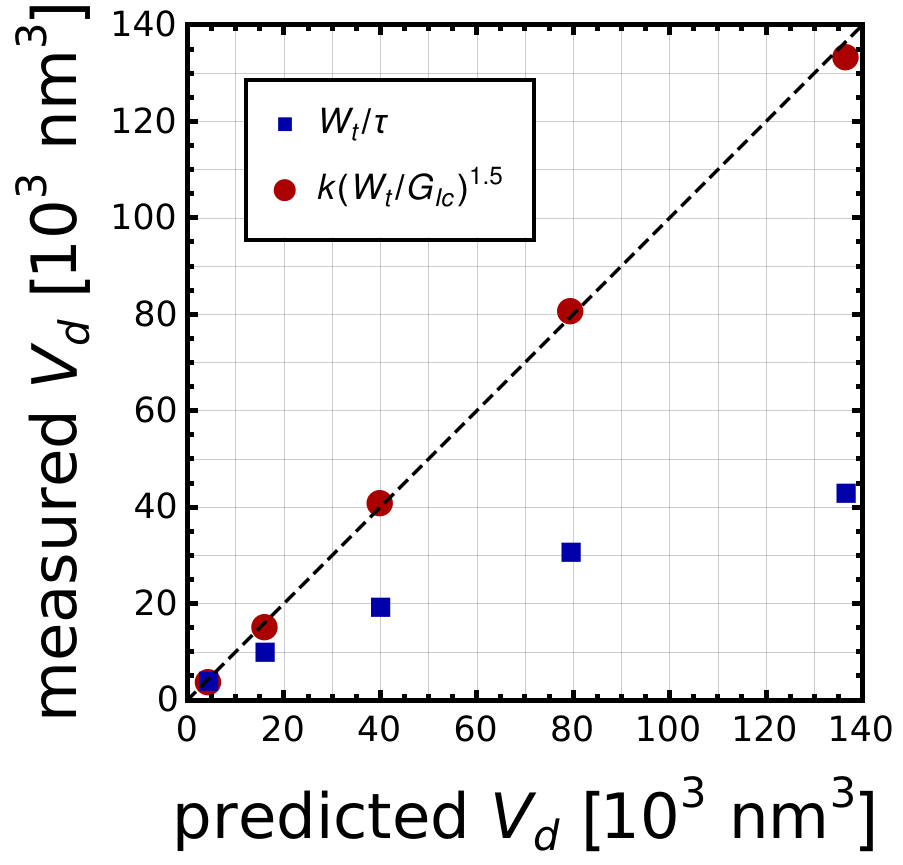}
\caption{\Large (a) prediction for $
\theta = 0^\text{o}$}\label{fig:6a}
\end{subfigure}
\begin{subfigure}[b]{.45\linewidth}
\includegraphics[width=\linewidth]{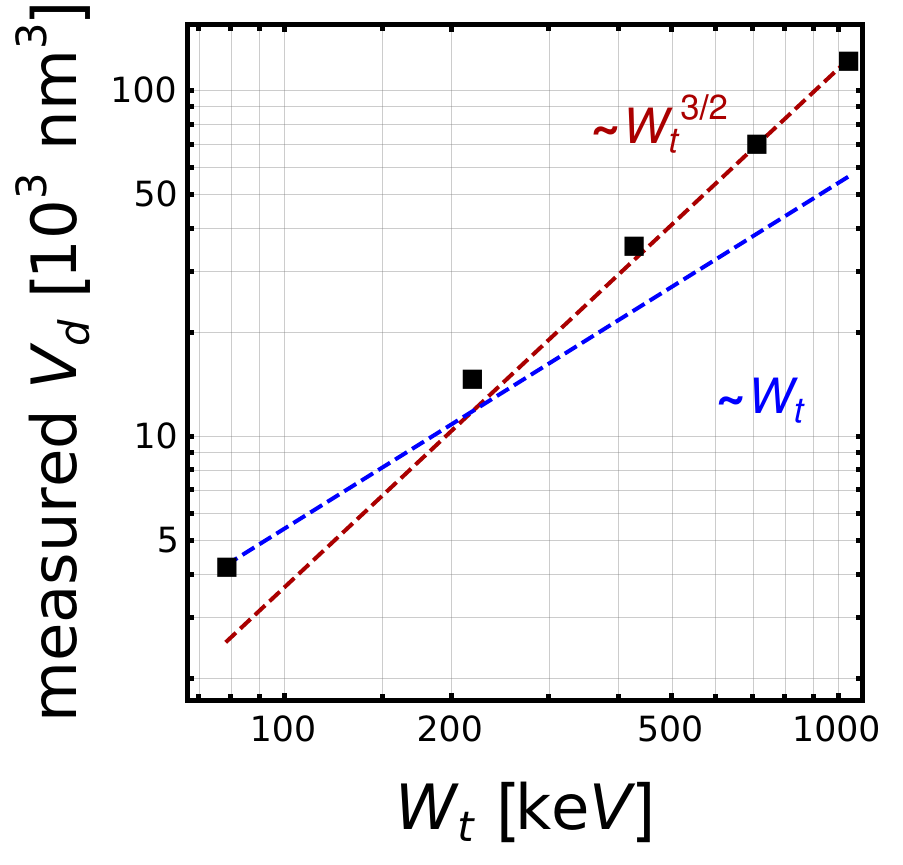}
\caption{\Large (b) scaling for $
\theta = 0^\text{o}$}\label{fig:6b}
\end{subfigure}
\begin{subfigure}[b]{.45\linewidth}
\includegraphics[width=\linewidth]{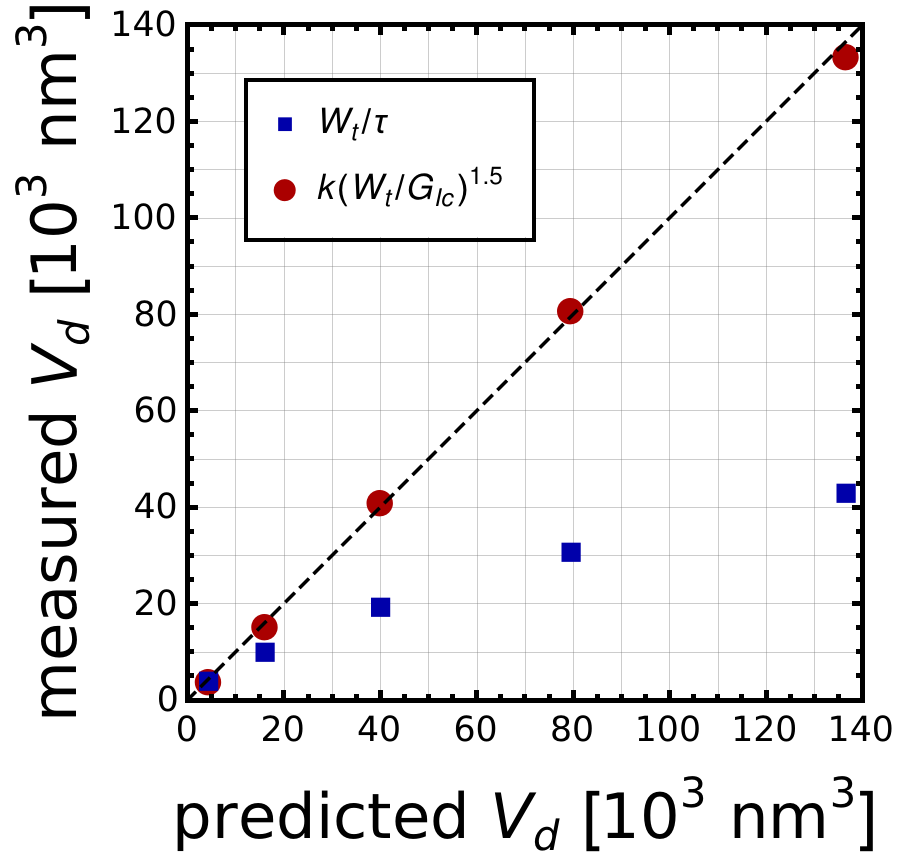}
\caption{\Large (c) prediction for $
\theta = 30^\text{o}$}\label{fig:6c}
\end{subfigure}
\begin{subfigure}[b]{.45\linewidth}
\bigskip
\includegraphics[width=\linewidth]{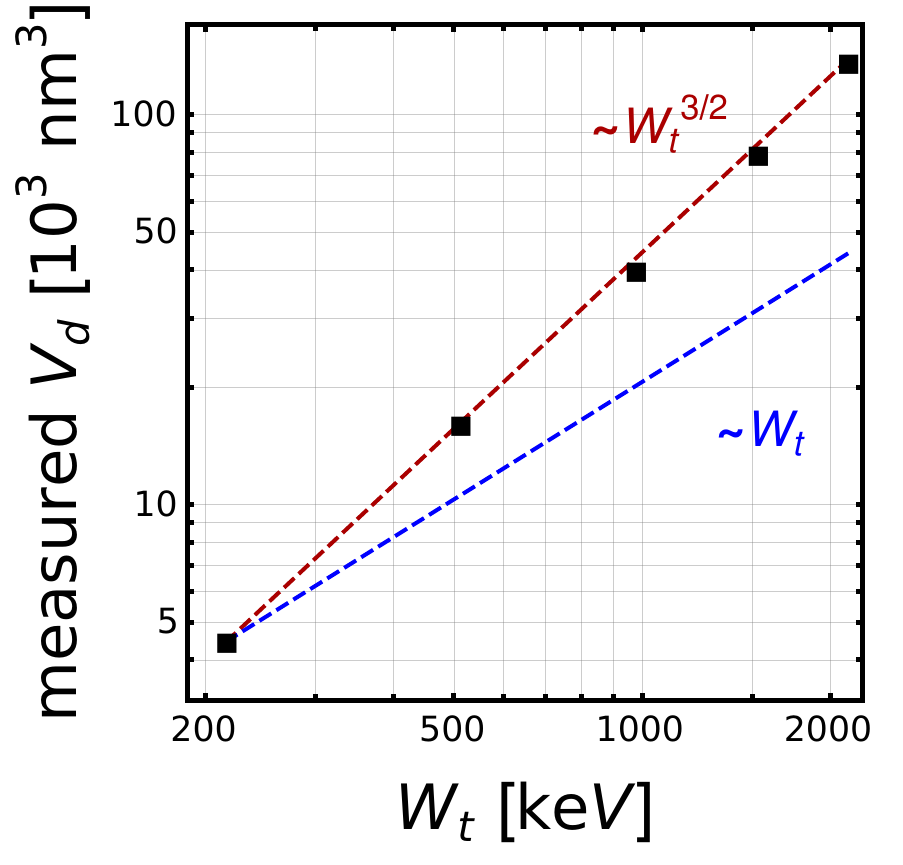}
\caption{\Large (d) scaling for $
\theta = 30^\text{o}$}\label{fig:6d}
\end{subfigure}
    \caption{{\color{black}
    Verifying the new scalings using 3D data.
    Panel (a): debris volume measured \textit{in silico} (y-axis) v. estimator (x-axis), 3D simulations with $\theta = 0^\text{o}$. The value of $\mathrm{k}$ is estimated from the crack pattern seen in the simulations, leaving $G_{Ic}$ as the only fit parameter; we found $\mathrm{k} = 0.285$ and $G_{Ic} = G_{Ic} = 29.0\,\mathrm{J/m^2} = 181\,\mathrm{eV/nm^2}$.
    Panel (b): data from 3D simulations with $\theta = 0^\text{o}$ in logarithmic scale alongside trend line for linear scaling ($\sim W_t$) and super-linear one ($\sim W_t^{1.5}$). Notice the intermediate transition zone from one to the other.
    Panel (c): similar to (a) for 3D simulations with $\theta = 30^\text{o}$. In this case $G_{Ic} = 61.2\,\mathrm{J/m^2} = 382\,\mathrm{eV/nm^2}$, the difference with respect to the other value being probably due to the different geometry-dependent crack pattern that is not accounted for.
    Panel (d): similar to (b) but for $\theta = 30^\text{o}$.
    }
    }
    \label{fig:scalings_3D}
\end{figure}

}

\section{Discussion}
\label{Sec:discussion}

{\color{black}
\Cref{fig:scalings_3D}(d) strongly substantiates the scalings presented in \cref{eq:new_scalings}(a). On the other hand, 
\Cref{fig:scalings_3D}(b) reveals a transition from a linear exponent to the super-linear one. It must be highlighted how the geometry (e.g., the angle $\theta$) affects the extent of the transition region. The super-linear scaling is evident for all simulations where $\theta = 30^{\text{o}}$, including those closer to $d^* \approx 18\, \mathrm{nm}$, whereas the transition does not happen until $d \approx 2d^*$ in the case $\theta = 0^{\text{o}}$. 
Hence, these results also give credence to the idea that the transition may depend on parameters as, e.g., the geometry (as this one controls the stress distribution).

We have been chiefly concerned with the qualitative trends in the tangential work -- debris volume relation, but we have also proposed new estimators \cref{eq:new_Vd_3D,eq:new_Vd_2D}. However, it remains to assess the fracture toughness $G_{Ic}$ via independent MD simulations to fully gauge their ability to predict numerical outcomes, instead of considering it a fit parameter. There is still no consensus as to how to estimate this material parameter using molecular dynamics but a number of options are currently being investigated \citep{Toughness_MD_1,Toughness_MD_2,Toughness_MD_3}.  
} 

We have shown in \cref{eq:scaling_d_dstar} that a linear relation between tangential work and the volume of worn material ($V_d \sim W_t$) can only hold when interlocking asperities fully undergo inelastic deformation, which in turn is only the case for the smallest asperities that lead to debris creation ($d > d^*$ while $d \approx d^*$).
Otherwise, we found $V_d \sim W_t^{2}$ in 2D (see appendix) and $V_d \sim W_t^{1.5}$ in 3D.
However, linearity between wear volume and work on the macroscopic scale has been reported in many experimental studies \citep{ref:rabinowicz1995friction,Burwell-Strang,Whittaker,Uetz,Fouvry}. For example, Archard's wear model \citep{Archard} states that
\begin{align} \label{eq:archard_equation}
    V_d 
    = 
    \mathrm{k}
    { F_n \cdot S 
        \over 
    \mathrm{H} } 
    \, ,
\end{align}
with $\mathrm{k}$ being the wear coefficient, $F_n$ the normal load, $S$ the sliding distance, and $\mathrm{H}$ the material hardness. So if $F_n$ can be related \textcolor{black}{linearly} to the tangential force $F_t$ (via any linear macroscopic friction law), then it follows that $V_d \sim F_t S \sim W_t$. 

A way to reconcile these facts is by accounting for the actual, statistical process of wear particle formation during sliding on a rough surface containing many asperity--asperity contacts. Contact simulations of rough surfaces combined with the application of \cref{eq:dstar} in a prior work \citep{Tobias_JMPS} have shown that most of the debris particles seem to arise from contacts close to $d^*$. Because the microcontacts grow from an initially small size, it stands to reason that their growth would be arrested by the emission of a wear particle when reaching $d = d^*$. In such a case it would follow that most of the worn mass stems from particles with sizes where the linearity is recovered. Note that an initially narrow distribution of particle sizes can later agglomerate into larger debris particles, giving rise to the plethora of sizes that can be seen in experiments (see, for instance, \citet{Son_experiments} and \citet{Weber}). It should be noted, though, that that model assumes that wear particles are formed at isolated contact spots. It has been proposed, however, that multiple, closely-spaced contacts can form a combined wear particle in an even more efficient process, although this is likely only the case under high normal load \citep{ramin_tobias_JF_PRL,Son}.

Even if the initial wear particle sizes were not restricted to the regime where linearity holds, it could still be the case that this non-linear micro-behavior gives rise to a linear relation as we upscale and more and more contacts are considered simultaneously (from asperities to clusters, from clusters to whole surfaces).
Such a phenomenon, i.e., non-linear interactions at the microscale resulting in a linear relation at the macroscale, is not unheard of in the field of tribology. For instance it is well known since Archard \citep{archard1957} and Greenwood and Williamson \citep{Greenwood1966} that the sum of Hertzian contacts with a random distribution of heights of contacting spheres---while strictly non-linear at the asperity level (that is the circular contact area is a non linear function of the local normal load)---gives rise to a linear dependence between real contact area and macroscopic normal load. 
This suggests that future research should focus on the collective behavior of asperity--asperity contacts on representative rough surfaces. 

\textcolor{black}{An additional, important aspect is the effect of loading rate on the FPZ under dynamic fracture conditions. It is well-known that the size of the FPZ depends on the rate, so the effect of loading rate over wear should also be treated in the framework presented here, for instance by adding correction terms. A preliminary study on the influence of this parameter is carried out in the appendix using idealized 2D simulations}. 

Finally, the possibility of other regimes where the volume--work scaling relation changes is not ruled out. 
We have ascertained that the fracture arrests before carving out the new particle completely, 
and that plastic hinges form in the ligaments left between the arrested crack tip and the free surface to finish the debris formation process (regard the strain distribution in \Cref{fig:Si_images} and the process depicted in \Cref{fig:process}). 
If the work that goes into plastifying this region scales proportionally to the volume of the asperities, 
then it may overcome the work associated to crack growth as the main contribution to the overall tangential work. Note that this process is geometry dependent, since the formation of the hinges occurs simultaneously with the crack propagation in the 3D simulations, suggesting that the scaling is not strongly affected in this specific case, at least. Techniques based on the phase field method \citep{Stella,Stella_2,Reinoso} seem ideal to further investigate such effects.

\section{Conclusion}
\label{Sec:conclusion}


We have analyzed three-dimensional molecular dynamics simulations, at a single asperity contact, resulting in the formation of adhesive wear particles. The simulations revealed a super-linear scaling of debris size with tangential work, contradicting a previous theoretical estimate with a linear scaling~\citep{PNAS}. A super-linear scaling could be observed because the simulations cover a wide range of contact junction sizes, reaching clearly above the critical junction size for debris formation \citep{critical_length_scale}. Our simulations indicate that for large contact junctions the process of debris creation is more efficient \textcolor{black}{than previously thought}, that is, less energy is required to form debris particles.

This has motivated the development of a new theoretical model, inspired by ductile fracture mechanics, as the process of debris formation is always accompanied by the propagation of cracks. The model accounts for two sources of energy dissipation, the new surfaces formed along the crack path area and a plastic volume around this crack path. The theory was shown \textcolor{black}{able to explain the trends observed in simulations.}

It is important to highlight that in the limit of contact junction size approaching the critical junction size, then the linearity between debris volume and tangential work is fully recovered. The rationale is that the critical junction size and the fracture process zone are essentially the same length scale. This results in a total plastification of the contact junction. 

Our findings provide a \textcolor{black}{roadmap towards a quantitative} framework to relate wear debris volume and frictional work and \textcolor{black}{should} be further informed with experimental observations. \textcolor{black}{ The next task would be to devise independent MD simulations to assess the fracture toughness.} A particular point of interest is that irrespectively of the asperity level mechanisms, a global linearity between total wear volume and frictional work is generally observed. This disconnection between asperity level mechanisms and global response requires further studies and would certainly be enriched by considering the sliding history.

\section*{Acknowledgements}

J. G.-S. and J.-F. M. gratefully acknowledge the support of the Swiss National Science Foundation (grant 200021\_197152, ``Wear across scales''). J. G.-S. would like to thank S. Z. Wattel for help with LAMMPS.

\section*{Supplementary material}

A \textit{Mathematica} notebook \citep{Mathematica} containing the computations leading to results (including figures) shown in the text is provided as Supplementary Material, and it can also be downloaded from the repository named \texttt{ductile\_wear} in the first author's GitHub page \texttt{github.com/jgarciasuarez}. 
All other materials necessary to reproduce results in this text can be obtained by correspondence to the authors.

\bibliography{references}  


\newpage 

\appendix
\renewcommand\thefigure{\thesection.\arabic{figure}}    

\section{Supplementary material: 2D simulations}
\setcounter{figure}{0}

\subsection{Numerical setting and results}

\begin{figure}[H]
\centering
\captionsetup[subfigure]{justification=centering}
\begin{subfigure}[b]{.45\linewidth}
\includegraphics[width=\linewidth]{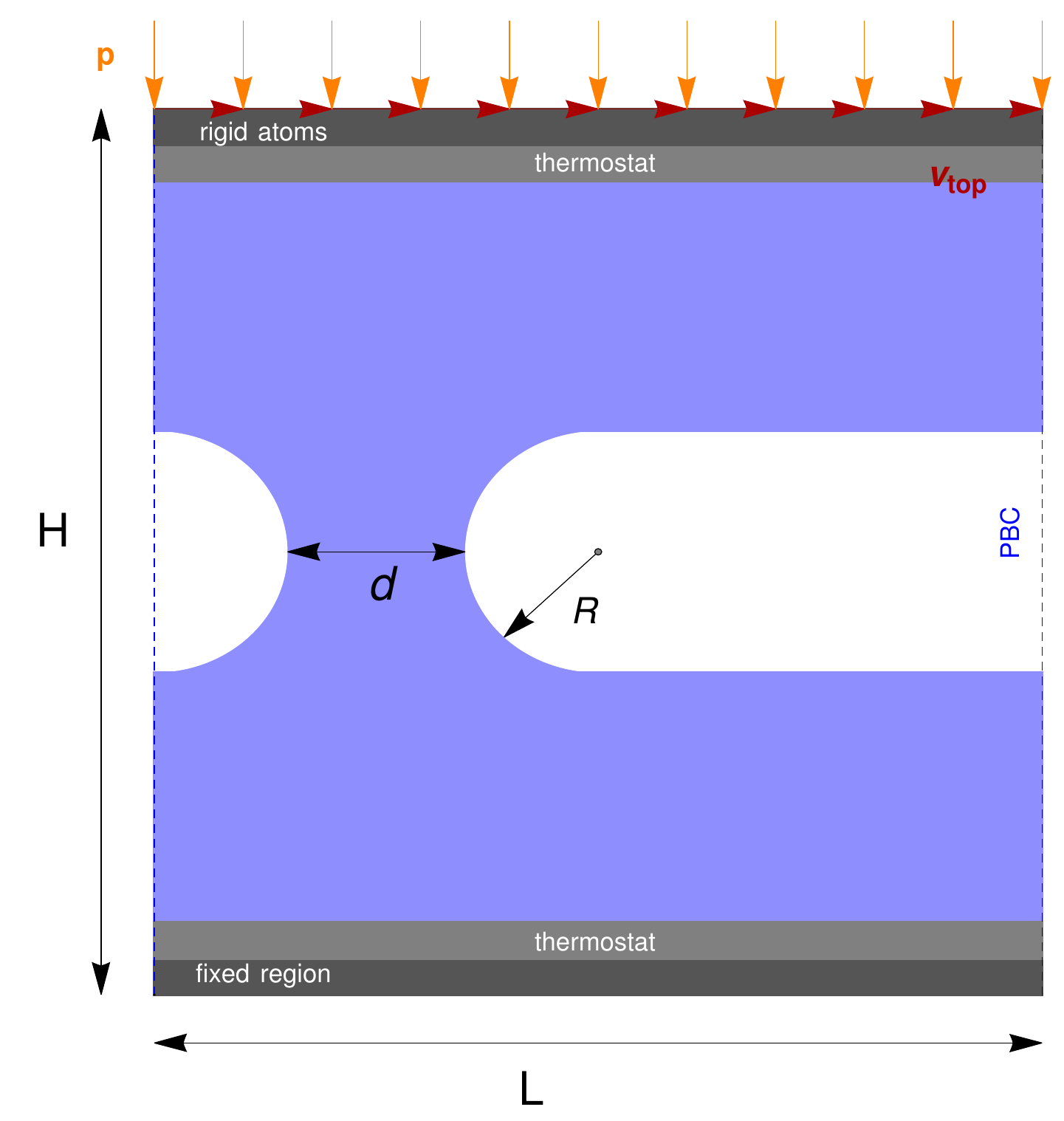}
\caption{\Large (a)}\label{fig:1c}
\end{subfigure}
\begin{subfigure}[b]{.45\linewidth}
\bigskip
\includegraphics[width=\linewidth]{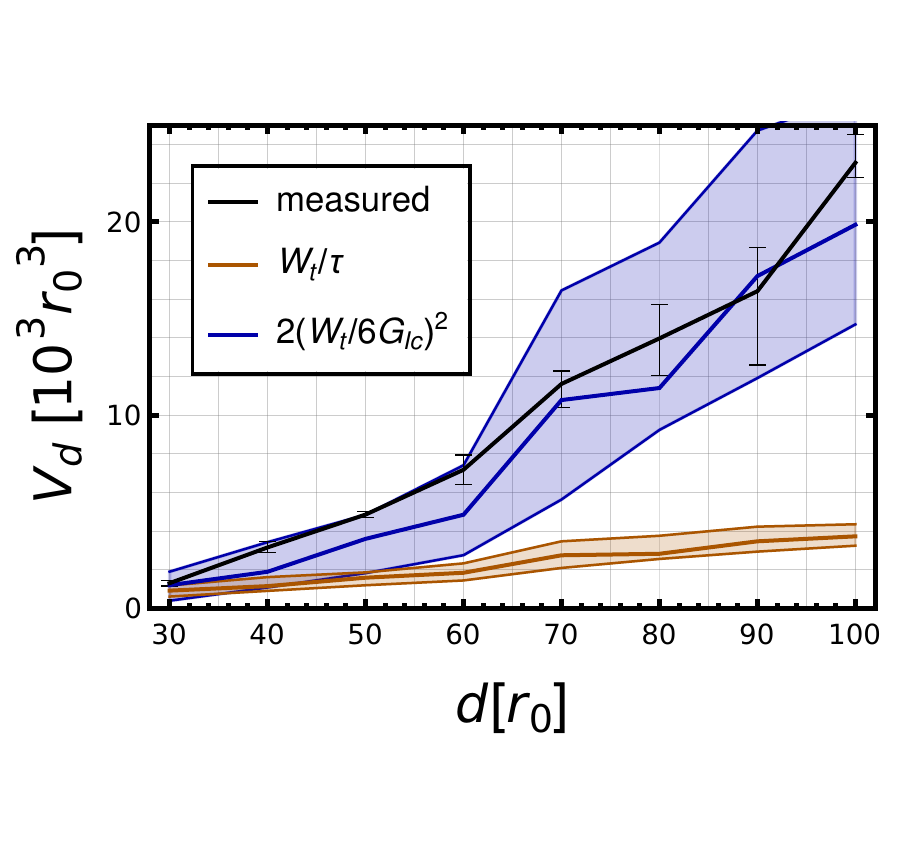}
\caption{\Large (b)}\label{fig:1d}
\end{subfigure}
    \caption{Evidencing the limitations of the debris volume estimate \cref{eq:Vdebris} \citep{PNAS} using both 2D simulations.
    Panel (a): scheme of the 2D simulations, including junction size and loading conditions. The geometry of the system is parametrized in terms of $d$, in particular $R=3d/4$, thus the gap between surfaces is $3/2d$. 
    \textcolor{black}{
    Panel (b): comparison of debris volume measured \textit{in silico} (black) v. estimate (orange) in 2D simulations , $v_{top} = 0.02 r_0/t_0$. The shaded orange region corresponds to maximum and minimum values of the estimates presented in \citep{critical_length_scale}. The shaded blue region correspond to maximum and minimum values of the \cref{eq:new_Vd_2D}.}
    }
    \label{fig:test_appendix}
\end{figure}

{\color{black}
Due to the large size of the simulations (on the order of 12\,million atoms), we were, however, not able to run many different geometries and could not run statistically independent repetitions for a given asperity size. Thus, we also used the computationally more affordable 2D simulations. Here, we could run a number of realizations of each case by initializing the atoms with different random velocity distributions consistent with the desired temperature.

These simulations hardly represent a practical situation (neither the simplified potential nor the 2D configuration are realistic), but they are not exempt of academic interest: in first place, they allow us to test the 2D estimator, \cref{eq:new_Vd_2D} and, moreover, the geometry in this case, \cref{fig:test_appendix}(a), features no sharp edges, so the lack of stress concentration leads to more variety in crack paths, what allow us to test the scaling in a disadvantageous situation. Finally, in this suite of virtual experiments we were able to repeat situations varying only the loading velocity $v_{top}$, so a first assessment of the result sensitivity to rate effects can be carried out. 


For each size, six realizations were run and the results are reported in \Cref{fig:test_appendix}(b) in black with error bars representing the range of values. The volume estimator (\cref{eq:Vdebris}) is shown in orange. It also covers a range of values indicated by the shaded region due to fluctuations of the total tangential work in the different realizations.

We used a modified Morse potential of the form \citep{Morse_Potential,critical_length_scale}

\begin{align}
    {V(r) \over \epsilon} =
    \begin{cases}
    (1-\exp{\{ -\alpha(r-r_0) \} })^2 - 1 
    &\text{\, if \,} 0 < r \le  1.1 r_0 \\
    {c_1 \over 6} r^3 + {c_2 \over 4} r^2 + {c_3 \over 6} r + c_4
    &\text{\, if \,} 1.1 r_0 < r \le  r_{cut} \\
    0 
    &\text{\, if \,}  r >  r_{cut}
    \end{cases}
    \, .
\end{align}

We used $\alpha = 3.93\,r_0^{-1}$ and expressed all data in reduced units of $\varepsilon$ and $r_0$. The cutoff distance was chosen as $r_{cut} = 1.422\,r_0$ and the parameters $c_1,\, c_2 , \, c_3, \, c_4$ were selected to ensure continuous energy and force at $r = 1.1\,r_0$, as well as zero energy and force at the cutoff.
This potential was used in \citet{critical_length_scale} and labeled ``P6''. 
The timestep is chosen to be $\Delta t = 0.005 t_0 = 0.005\,r_0/\sqrt{\varepsilon/m}$ (where $m$ is the atomic mass).

We then used atoms on a hexagonal lattice and cut out the geometry sketched in Fig.~\ref{fig:test}(c). 
The height of the interlocking asperity is $2R = 3d/2$. 
The box size is chosen proportionally to $d$: its total height $H$ equals $8d$, while its horizontal length is $L=5d$. These values are chosen to guarantee that the model's horizontal edges and the periodic boundary conditions (PBCs) on the vertical ones do not affect the local stress state at the junction. Langevin thermostats are located on top and bottom layers to enforce a constant temperature $T=0.025\,\epsilon/k_B$, where $k_B$ represents the Boltzmann constant. 
The bottom layer is constrained to remain fixed, while a normal pressure of $0.03 \, \epsilon/r_0^3$ and a sliding velocity are applied to the top one.
These simulations are run over a minimum sliding distance equal to $3d$, to ensure enough sliding so as to trigger debris creation. 
The sliding velocity is $0.02\,r_0/t_0$, additional simulations that were run at velocity $0.05\,r_0/t_0$ are also reported. 
The pre-formed junction sizes considered in these simulations range from $d=30\,r_0$ to $d=100\,r_0$ in increments of $10\,r_0$. The minimal value of $d$ that yields debris creation is $d^* \approx 15\,r_0$, meaning all these simulations lead to third-body formation.
The number of atoms in the largest box is $\approx 40600$ which makes these 2D simulations relatively affordable, thereby allowing six independent realizations of each size for statistics.

We estimate the debris volume (area in 2D), by counting the number of atoms in the debris particle and multiplying them by the average atomic volume (area) in the bulk. 

We also estimate the shear strength via independent simulations, finding $\tau = 0.78\,\epsilon / r_0^3$.
}

\begin{figure}
    \centering
    \includegraphics[width=0.95\textwidth]{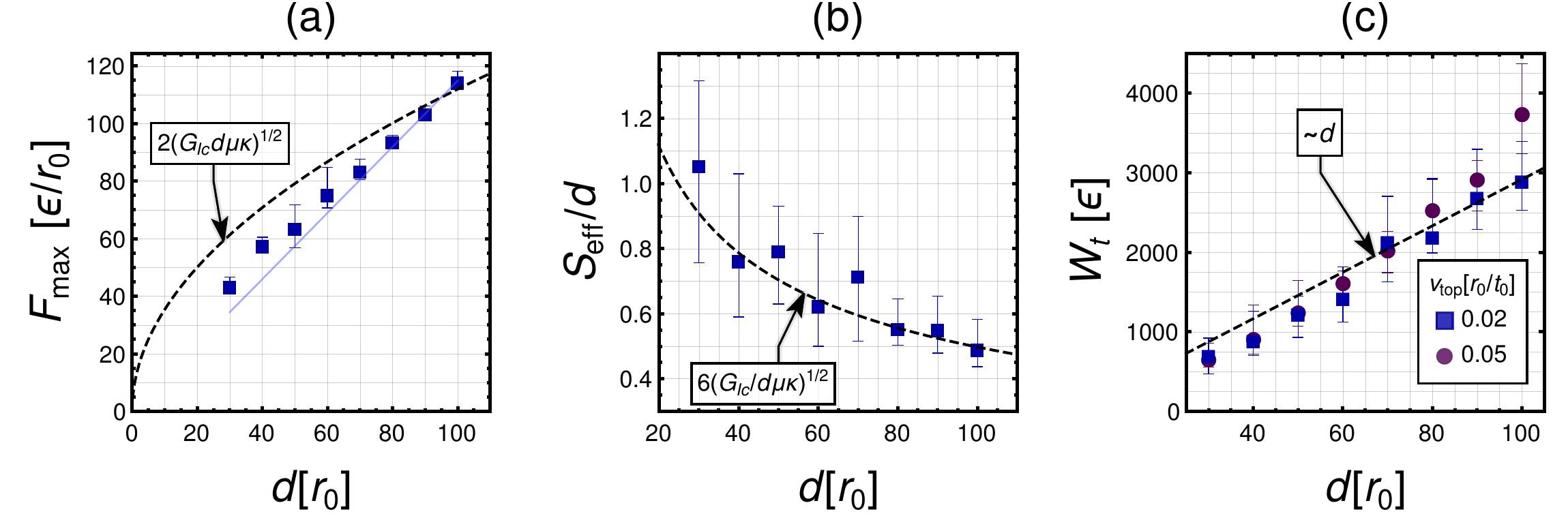}
    \caption{\textcolor{black}{Verifying the assumptions behind \cref{eq:Vdebris}. (a) Assumption \#1: maximum force. Error bars represent the range of values obtained in numerical simulations ($v_{top} = 0.02 r_0 / t_0$), dashed line represents the values from Continuum Mechanics, solid blue line represents the linear trend defined by the last two points. $G_{Ic}$ in our case represents a fit parameter once we fix an assumed total crack length ($l_{crack} = 6 d$, see \Cref{fig:process}), its value $4.86\, \epsilon/r_0^2$ is inferred directly from simulations by fitting the linear data presented in panel(c). (b) Assumption \#2: effective sliding distance. Error bars represent the range of values obtained in numerical simulations ($v_{top} = 0.02 r_0 / t_0$), the dashed line represents the values from Continuum Mechanics. (c) Work scaling: linear increase in tangential work necessary to generate debris, $v_{top} = 0.02 r_0 / t_0$ (squares) and $v_{top} = 0.02 r_0 / t_0$ (circles), notice apparent rate effects (deviation from linear trend) for largest size and larger velocity.}}
    \label{fig:assumptions_check}
\end{figure}

{ \color{black}
Taking advantage of the fact that the height that appears in \cref{eq:S_eff} can be written as $h = 3/4 d$, \Cref{fig:test}(c),
\Cref{fig:assumptions_check} panel (a) depicts the numerical results for $S_{eff}/d$ from the 2D model in \Cref{fig:test}(a) concurrently with a scaling trend consistent with \cref{eq:S_eff}. 
Note that the prefactor is adapted slightly to account for the shape of the asperities in the simulation not being prismatic.

Even though the 2D results we have obtained agree well with the predicted trends, we remark that for the largest size we tested, $d=100 \, r_0$, there is a difference between the results for the two velocities considered, Figure4(a). This hints to a growing sensitivity to loading rate as the asperity size increases.
}

\newpage

\subsection{Example of fracture process}

\begin{figure}[H]
    \centering
    \includegraphics[width=0.9\textwidth]{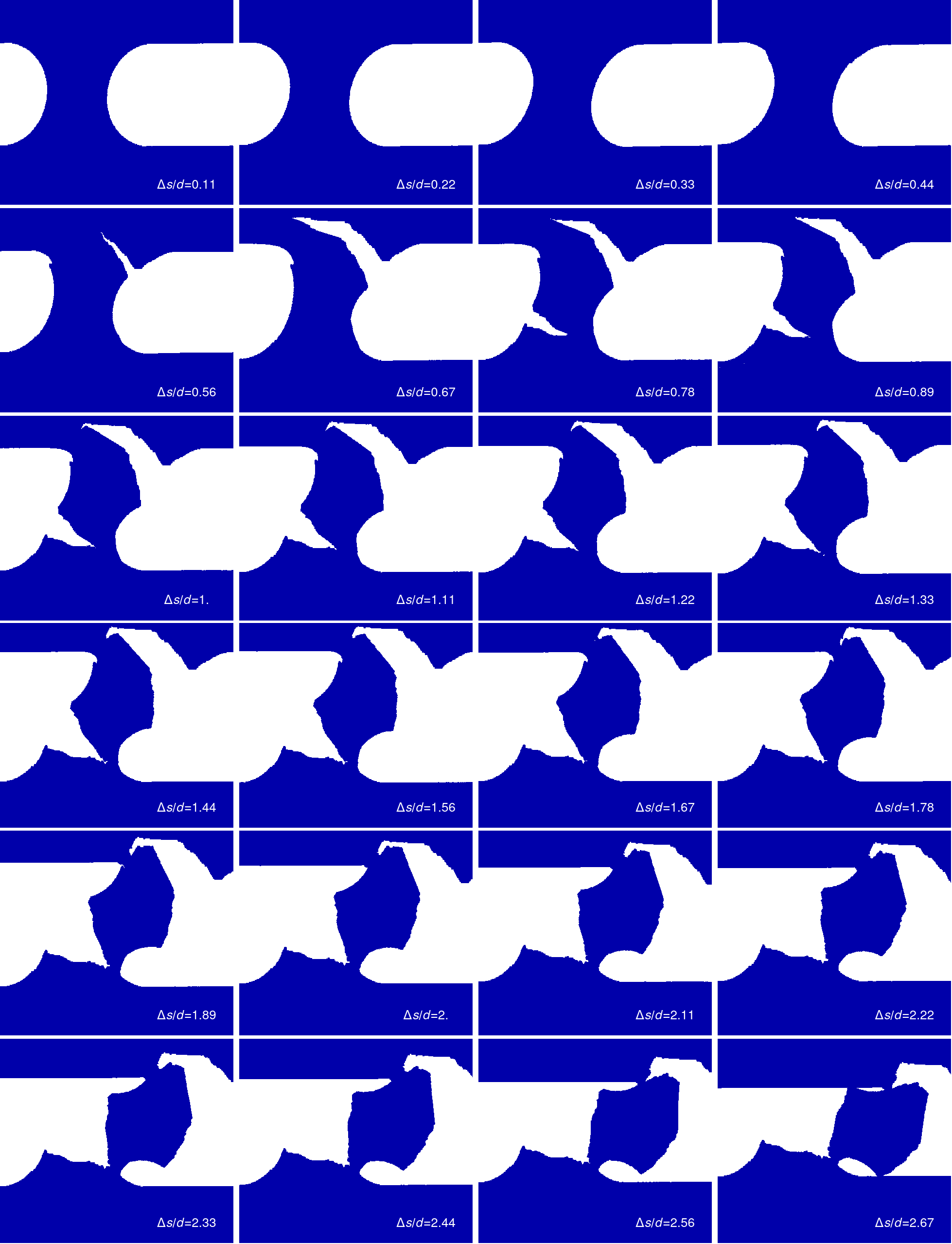}
    \caption{Snapshots of the third-body formation process (for size $d=90\,r_0$), images include cumulative dimensionless sliding $\Delta s / d$. Process runs from left to right, from top to bottom.}
    \label{fig:process}
\end{figure}

\section{Further evidence of propagation regimes}

\begin{figure}[H]
    \centering
    \includegraphics[width=\textwidth]{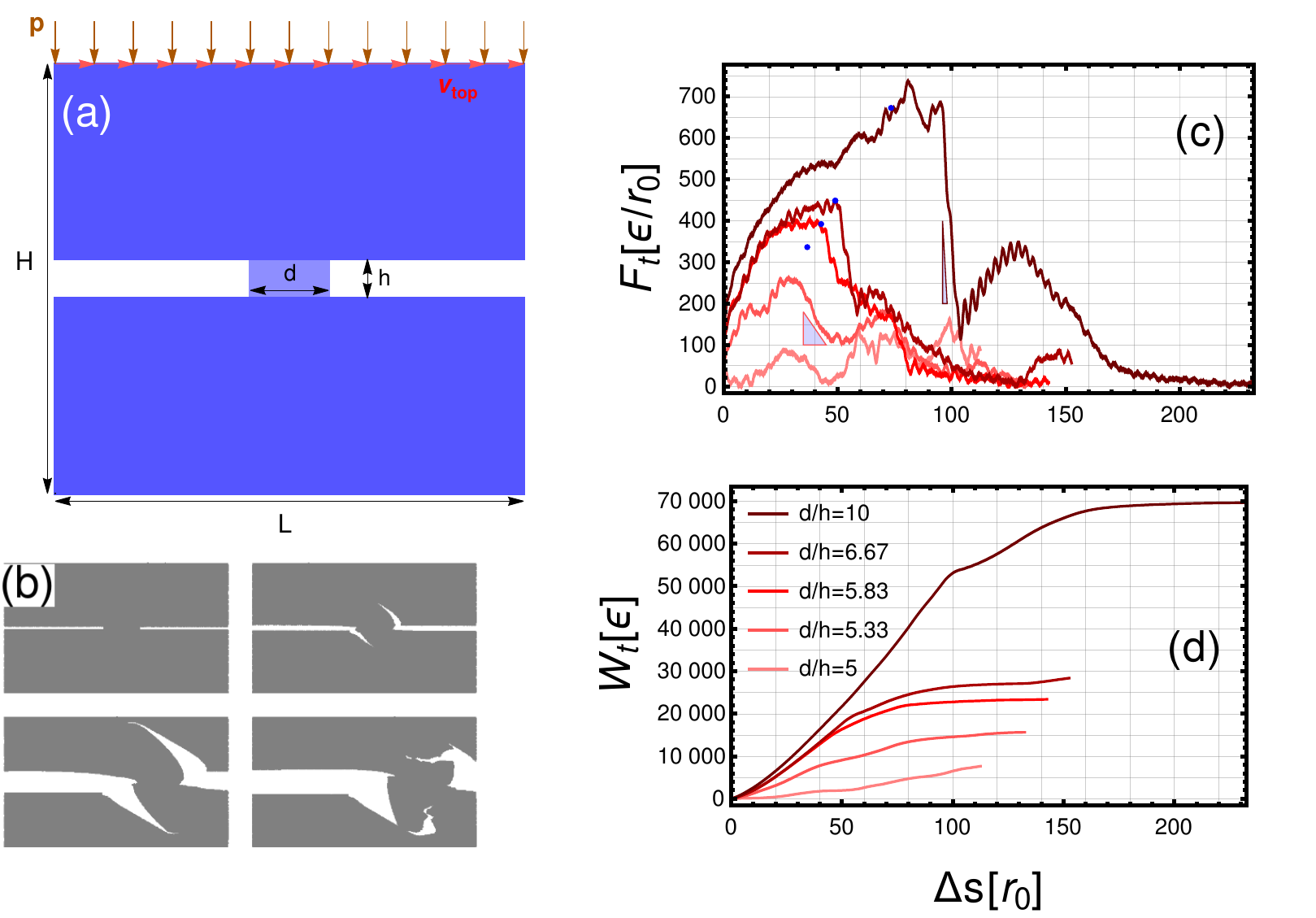}
    \caption{Influence of crack propagation during wear. (a) Scheme of the displacement-controlled system with ``handshake'' pre-formed junction (as presented in \citet{Son}), all junctions $h=6r_0$. (b) Snapshots of crack propagation leading to debris detachment. (c) Tangential force evolution for different junction sizes, the blue triangles highlight the slope of the unloading in the case $d/h=5.33$ (slow unloading proportional to imposed velocity) and $d/h=10$ (fast unloading proportional to wave velocity). (d) Tangential work evolution for different sizes.}
    \label{fig:Son_data}
\end{figure}

This study uses quasi-2D simulations, with atoms arranged on an FCC lattice, featuring rectangular pre-formed junctions \citep{Son} (the trajectory of the asperities prior to contact was neglected). All units are expressed in the corresponding potential reduced units, similarly to the 2D simulations presented in the body of the article, see \Cref{Sec:methods}. These correspond to displacement-controlled virtual experiments, where the top surface sliding velocity is $0.01 r_0/t_0$. 
Five different junction widths $d$ were considered, $d/r_0 = 30, \, 32, \, 35, \, 40, \, 60$ ($r_0$ is the distance between the atom center and the potential well), while fixing the junction height $h / r_0 = 6$, thus $d/h = 5, \, 5.33, \, 5.83, \, 6.67, \, 10$. 
All simulations run until the total sliding distance reaches $50r_0$.  
The critical junction size was $d^* \approx 35 r_0$.
All the other details can be checked in \citet{Son}. 

The evolution of the tangential force for each size is visualized in \Cref{fig:Son_data}(c). 
The qualitative dissimilarities are evident. 
In the lightest color, the smallest size, $d/r_0=30$ and $d/h=5$, was chosen so as not to generate debris (plastic asperity smoothing, since $d^* \approx 31 r_0$). All the others do lead to particle separation (notice the different convexity of the work evolutions in \Cref{fig:Son_data}(c)). 

See that, as the junction width increases, gradients in the force profile become acuter; for the two largest sizes, there is a sudden drop, which entails drastic loss of stiffness and fast energy release, coherent with unstable crack propagation. 
The following bump represents work invested not in separation, but in rotating the already-formed particle out of the hollow left behind by the atoms that now belong to the contour of the debris. 

The qualitative change evidences the transition between the strength-controlled regime and the toughness-controlled one.

\section{Snapshots from 3D simulations for $\theta = 0^\circ$}

\begin{figure}[H]
\centering
  \includegraphics[width=\linewidth]{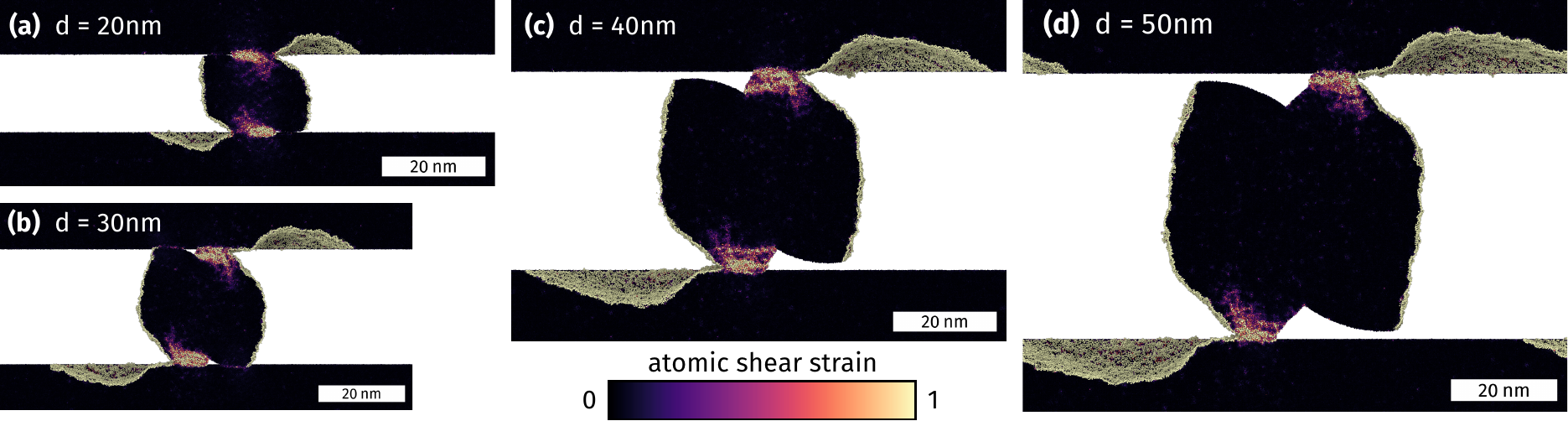}
  \caption{Plastic strain for the 3D simulations with $\theta = 0^\circ$. A slice through the middle of the wear particle is shown to visualize the plasticity in the bulk. (a) For the smallest asperity diameter close to $d^* = 18\,\mathrm{nm}$, significant plasticity occurs in the bulk of the particle. (b)--(d) With increasing asperity size, the plasticity becomes more localized and does not fill the whole particle.}
\end{figure}

\end{document}